\documentclass[aps,prl,twocolumn,floats,nofootinbib,longbibliography]{revtex4-1}

\usepackage{graphicx}
\usepackage{graphics}
\graphicspath{{./figs/}}

\usepackage{bm}
\usepackage{amsmath,amssymb}
\usepackage{aas_macros}
\usepackage{natbib}
\usepackage[colorlinks=true]{hyperref}
\usepackage{breqn}

\usepackage{color}
\newcommand{\colp}[1]{\textcolor{black}{#1}}

\def\aap{\textit{Astronomy and Astrophysics}}

\def\be{\begin{equation}}
\def\ee{\end{equation}}
\def\bea{\begin{eqnarray}}
\def\eea{\end{eqnarray}}

\def\lg{\lambda_g}
\newcommand\Ol{\mathcal O}
\begin{document}
\title{Constraining the mass of the graviton with the planetary ephemeris INPOP}

\author{
L. Bernus $^{1}$,
O. Minazzoli $^{3,4}$,
A. Fienga $^{2,1}$,
M. Gastineau $^{1}$,
J. Laskar $^{1}$,
P. Deram $^2$\\
$^{1}$IMCCE, Observatoire de Paris, PSL University, CNRS, Sorbonne Universit\'e, 77 avenue Denfert-Rochereau, 75014 Paris, France \\
$^{2}$G\'eoazur, Observatoire de la C\^ote d'Azur, Universit\'e C\^ote d'Azur, IRD, 250 Rue Albert Einstein, 06560 Valbonne, France\\
$^{3}$Centre Scientifique de Monaco, 8 Quai Antoine 1er, Monaco \\
$^{4}$Artemis, Universit\'e C\^ote d'Azur, CNRS, Observatoire de la C\^ote d'Azur, BP4229, 06304, Nice Cedex 4, France
}

\begin{abstract}
We use the planetary ephemeris INPOP17b to constrain the mass of the graviton in the Newtonian limit. We also give an interpretation of this result for a specific case of fifth force framework. We find that the residuals for the Cassini spacecraft significantly (\colp{90}\% C.L.) degrade for Compton wavelengths of the graviton smaller than \colp{$1.83\times 10^{13}$} km, corresponding to a graviton mass bigger than \colp{$6.76\times 10^{-23} eV/c^2$}. This limit is comparable in magnitude to the one obtained by the LIGO-Virgo collaboration in the radiative regime. We also use this specific example to illustrate why constraints on alternative theories of gravity obtained from postfit residuals are generically overestimated.
\end{abstract}
\maketitle

\section{Introduction}

From a particle physics point of view, general relativity can be thought as a theory of a massless spin-2 particle --- hereafter named \textit{graviton}. From this perspective, it is legitimate to investigate whether or not the graviton could actually possess a mass --- even if minute. Such an eventuality has been scrutinized from a theoretical point of view since the late thirties, with the pioneer work of Fierz and Pauli \cite{fierz:1939rs}. There is a wide set of massive gravity theories, which lead to various phenomenologies \cite{derham:2017rp}. One of the generic prediction from several models --- although not all of them \footnote{In particular, usually not for models prone to the Vainshtein mechanism \cite{derham:2017rp}.} --- is that the usual $1/r$ falloff of the Newtonian potential acquires a Yukawa suppression \cite{derham:2017rp}. In the present manuscript, we aim to test this particular phenomenology, regardless of the specificity of the theoretical model that produced it. For more information on the status of current theoretical models, we refer the reader to \cite{derham:2014lr,derham:2017rp}. As a consequence, we assume that the line element in a space-time curved by a spherical massive object at rest, at leading order in the Newtonian regime, reads

\bea
ds^2 &=& \left( -1 + \frac{2GM}{c^2 R}~e^{-R/\lg} \right) c^2 dT^2 \nonumber \\
&&+ \left( 1 + \frac{2GM}{c^2 R}~e^{-R/\lg} \right) dL^2, \label{eq:visserLE}
\eea
with $dL^2 \equiv dX^2+dY^2+dZ^2$, $R\equiv\sqrt{X^2+Y^2+Z^2}$ and $\lg$ the Compton wavelength of the graviton --- although we will see that our constraints can be applied on a wider range of massive and non-massive gravity metrics.
Obviously, as long as $\lg$ is big enough, the gravitational phenomenology in the Newtonian regime can reduce to the one of general relativity to any given level of accuracy.

Another generic feature of many massive gravity theories is that, if the graviton is massive, its dispersion relation may be modified according to $E^2 = p^2c^2 + m_g^2 c^4$ \cite{derham:2017rp}, such that the speed of a gravitational waves depends on its energy (or frequency) $v_g^2/c^2 = c^2 p^2 / E^2 \simeq 1 - h^2 c^2/(\lg^2 E^2)$. Therefore, the waveform of gravitational waves would be modified during their propagation, while at the same time, sources of gravitational waves have been seen up to more than $1420$\,Mpc (at the 90\% C.L.) \cite{ligo:2018ax}. 
As a consequence, waveform match filtering can be used to constrain the graviton mass from gravitational waves detections \cite{will:1998pr,delpozzo:2011pr}. 

Combining bounds from several events in the catalog GWTC-1 \cite{ligo:2018ax} leads to $\lg \geq 2.6 \times 10^{13}$\,km (resp. $m_g \leq 5.0 \times 10^{-23} $eV/$c^2$ \cite{ligo:2018ax,ligo:2019ar} \footnote{With the definition $m_g = h / (c\lg)$.}) at the 90\% C.L \footnote{Assuming that the graviton mass affects the propagation only, and not the binaries dynamics.}. It is important to keep in mind that this limit is obtained in the radiative regime, while we focus here on the Newtonian regime. Although, one could expect $\lg$ to have the same value in both regimes for most massive gravity theories, it may not be true for all massive gravity theories. Therefore, both constraints should be considered independently from an agnostic point of view. See, e.g., \cite{derham:2017rp} for a review on the graviton mass constraints.

\section{Importance of a global fit analysis}

Twenty years ago \cite{will:1998pr} and more recently \cite{will:2018cq}, Will argued that Solar System observations could be used to improve --- or at least be comparable with --- the constraints on $\lg$ obtained from the LIGO-Virgo Collaboration --- assuming that the parameters $\lg$ appearing in both the radiative and Newtonian limits are the same. A graviton mass would indeed lead to a modification of the perihelion advance of Solar System bodies. Hence, based on current constraints on the perihelion advance of Mars --- or on the post-Newtonian parameters $\gamma$ and $\beta$ --- derived from Mars Reconnaissance Orbiter (MRO) data, Will estimates that the graviton's Compton wavelength should be bigger than $(\colp{1.2 - 2.2}) \times 10^{14}$km (resp. $m_g < (\colp{5.6-10}) \times 10^{-24}$ eV/$c^2$), depending on the specific analysis. 
However, as an input for his analysis, Will uses results based on the statistics of residuals of the Solar System ephemerides that are performed without including the effect of a massive graviton. But various parameters of the ephemeris (eg. masses, semi-major axes, Compton parameter, etc.) are all more or less correlated to $\lg$ (see Table \ref{table:corr}). Therefore, any signal introduced by $\lambda_g < + \infty$  --- for instance, a modification of a perihelion advance --- can in part be re-absorbed during the fit of other parameters that are correlated with the mass of the graviton. (See Supplemental Materials).
This necessarily leads to a decrease of the constraining power of the ephemeris on the graviton mass with respect to postfit estimates. As a corollary, all analyses based solely on postfit residuals tend to overestimate the constraints on alternative theories of gravity due to the lack of information on the correlations between the various parameters. 
Eventually, one cannot produce conservative estimates of any parameter without going through the whole procedure of integrating the equations of motion and fitting the parameters with respect to actual observations --- which is the very \textit{raison d'\^etre} of the ephemeris INPOP.

\begin{table}
\resizebox{\columnwidth}{!}{
\begin{tabular}{c|ccccccc}
		~ &$\lambda_g$& $a$ Mercury  &$a$ Mars & $a$ Saturn& $a$ Venus &$a$ EMB & $GM_{\odot}$ \\
		\hline
		$\lambda_g$ &1 &0.50&0.49&0.04&0.39&0.05&0.66\\
        $a$ Mercure &$\cdots$& 1& 0.21& 0.001& 0.97& 0.82& 0.96 \\
		$a$ Mars  &$\cdots$& $\cdots$& 1 &0.03&0.29& 0.53 &0.06 \\
		$a$ Saturn &$\cdots$& $\cdots$ &$\cdots$& 1& 0.003 & 0.02& 0.01 \\
		$a$ Venus &$\cdots$&$\cdots$&$\cdots$&$\cdots$&1& 0.86& 0.94\\
		$a$ EMB &$\cdots$&$\cdots$& $\cdots$& $\cdots$ &$\cdots$&1 &0.73\\
		$GM_{\odot}$  &$\cdots$&$\cdots$ &$\cdots$& $\cdots$&$\cdots$&$\cdots$ &1\\
	\end{tabular}
}
	\caption{Examples of correlations between various INPOP17b parameters and the Compton wavelenght $\lg$. $a$, EMB and $M_{\odot}$ state for semi-major axes, the Earth-Moon barycenter and the mass of the Sun respectively.}
	\label{table:corr}
\end{table}

INPOP (Int\'egrateur Num\'erique Plan\'etaire de l'Observatoire de Paris) \cite{fienga:2008aa} is a planetary ephemeris that is built by integrating numerically the equations of motion of the Solar System following the formulation of \cite{moyer:2003}, and by adjusting to Solar System observations such as lunar laser ranging or space missions observations. In addition to adjusting the astronomical intrinsic parameters, it can be used to adjust parameters that encode deviations from general relativity \cite{fienga:2011cm,verma:2014aa,fienga:2015cm,viswanathan:2018mn}, such as $\lg$. The latest released version of INPOP, INPOP17a \cite{viswanathan:2018dc}, benefits of an improved modeling of the Earth-Moon system, as well as an update of the observational sample used for the fit \cite{viswanathan:2018mn} --- especially including the latest Mars orbiter data. For this work we use an extension of INPOP17a, called INPOP17b, fitted over an extended sample of Messenger data up to the end of the mission, provided by \cite{2016JGRE..121.1627V}.

In the present communication, our goal is to use the latest planetary ephemeris INPOP17b in order to constrain a hypothetical graviton mass directly at the level of the numerical integration of the equations of motion and the resulting adjusting procedure. By doing so, the various correlations between the parameters are intrinsically taken into account, such that we can deliver a conservative constraint on the graviton mass from Solar System obervations --- details about the global adjusting procedure are given in Supplemental Material.

\section{Modelisation for Solar System phenomenology}

Following Will \cite{will:2018cq}, we develop perturbatively the potential in terms of $r/\lg$, such that the line element (\ref{eq:visserLE}) now reads

\bea \label{eq:ds2}
&&ds^2 = \left( -1 + \frac{2GM}{c^2 r}\left[1+ \frac{1}{2} \frac{r^2}{\lg^2} \right] \right) c^2 dt^2 \\
&&+ \left( 1 + \frac{2GM}{c^2 r}\left[1+ \frac{1}{2} \frac{r^2}{\lg^2} \right] \right) dl^2+ \Ol(c^{-3}\lg^{-2}), \nonumber
\eea
albeit with a change of coordinate system \footnote{We assume that the underlying theory of gravity is covariant, such that this change of coordinates has no impact on the derivation of the actual \textit{observables}.}
\begin{equation}\label{eq:change}
	T = \frac{t}{\sqrt{1+ \frac{GM}{c^2 \lg}}}, \quad X^i = \frac{x^i}{\sqrt{1- \frac{GM}{c^2 \lg}}}
\end{equation}
The change of coordinate system is meant to get rid of the non-observable constant terms that appear in the line element Eq. (\ref{eq:visserLE}) after expanding in terms of $\lg^{-1}$. Considering a N-body system, the resulting additional acceleration to incorporate in INPOP's code is
\be
\delta a^i = \frac{1}{2} \sum_P \frac{G M_P}{\lg^2} \frac{x^i-x_P^i}{r}+\Ol(\lg^{-3}), \label{eq:accgravmas}
\ee
where $M_P$ and $x_P^i$ are respectively the mass and the position of the gravitational source $P$. In what follows, we make the standard assumption that the underlying theory is such that light propagates along null geodesics \cite{derham:2014lr}.
From the null condition $ds^2 =0$ and Eq. (\ref{eq:ds2}), the resulting additional Shapiro delay at the perturbative level reads

\bea
\delta T_{ER} &=&\frac{1}{2} \sum_P \frac{GM_P}{c^3 \lg^2} \left[{\vec N}_{ER} \cdot \left({\vec R}_{PR} R_{PR} - {\vec R}_{PE} R_{PE} \right) \right. \\
&+&\left.  b_{P}^2 \ln \left(\frac{R_{PR} + {\vec R}_{PR} \cdot  {\vec N}_{ER}}{R_{PE} + {\vec R}_{PE} \cdot  {\vec N}_{ER}} \right) \right]+ \Ol(c^{-3}\lg^{-3}), \nonumber
\eea
where ${\vec R}_{XY} = {\vec x}_Y - {\vec x}_X$, $R_{XY} = |{\vec R}_{XY}|$, ${\vec N}_{XY} = {\vec R}_{XY}/R_{XY} $ and  $b_P = \sqrt{R_{PE}^2 - ({\vec R}_{PE} \cdot {\vec N}_{ER})^2}$. One can notice that the correction to the Shapiro delay scales as $(L_c/\lg)^2$ with respect to the usual delay, where $L_c$ is a characteristic distance of a given geometrical configuration. Given the old acknowledged constraint from Solar System observations on the graviton mass ($\lg > 2.8 \times 10^{12}$\,km \cite{will:2018cq,talmadge:1988pl,will:1998pr}), one deduces that the correction from the Yukawa potential on the Shapiro delay is negligible for past, current and forthcoming radio-science observations in the Solar System \footnote{However, note that the scaling of the correction to the Shapiro delay illustrates the breakdown of the $\lg^{-1}$ development in cases where the characteristic distances involved are large with respect to the Compton wavelength --- as it should be expected.}.

On the other hand, the fifth force formalism predicts an additional Yukawa term to the Newtonian potential \cite{hees:2017prl}
\begin{equation}
	V = \frac{Gm}{r}(1+\alpha e^{-r/\lambda})
\end{equation}
If we assume that $\lambda\gg r$ and $\alpha>0$, we can also expand the Yukawa term, such that our result on $\lg$ can be transposed to $\lambda/\sqrt{\alpha}$ --- although, one first has to rescale the gravitational constant to $\tilde{G}=G(1+\alpha)$, and then to make the same coordinates change as in Eq. \eqref{eq:change}, but substituting $\lg$ by $\lambda/\alpha$. 
Note that a fifth force is also one of the generic features of several massive gravity theories \cite{derham:2017rp}.

\section{Evaluation of the significance of the residuals deterioration}

To give a confidence interval for $\lg$, we proceed as follows. For each value of $\lg$, we perform a global fit of all other parameters to observations using the same data that for the reference solution INPOP17b --- therefore, for the same number of observations.
After the global fit procedure, we compute the residuals at the same dates that for the reference solutions and look how they are degraded or improved with respect to $\lg$. The result is that Cassini residuals are the first to degrade significantly while $\lg$ decreases (see Supplemental Material for details).

To quantify the statistical meaning of this degradation, we perform a Pearson \cite{Pearson:1992} $\chi^2$ test between both residuals in order to look at the probability that they were both built from the same distribution. To compute the $\chi^2$, we build an optimal histogram with the Cassini residuals of INPOP17b using the method described in \cite{scott:1979bm}, assuming the gaussianity of the distribution of the residuals. We determine the optimal bins in which are counted the residuals to build the histogram. Then, using the same bins, we build an histogram for the Cassini residuals obtained by the solution to be tested with a given value of $\lg$. Note that the first bin left-borned is $-\infty$ and the last bin right-borned is $+\infty$. Let $(C_i)_i$ be the bins in which are counted the values of the residuals and $N_i^I$, $N_i^G$ be the number of residuals of INPOP17b and the solution to be tested, respectively, counted in bin number $i$. One can then compute
\begin{equation}
	\chi^2(\lg) = \sum_{i=1}^n \frac{(N_i^G - N_i^I)^2}{N_i^I} 	
\end{equation} 
For Cassini data, it occurs that the optimal binning gives 10 bins. As a result, this $\chi^2$ follows a $\chi^2$ law with 10 degrees of freedom. If the computed $\chi^2$  is then greater than its quantile for a given confidence probability $p$, we can say that the distribution of the residuals obtained for $\lg$ is different from the residuals obtained by the reference solution with a probability $p$. This test can be done for both a positive detection of a physical effect and a rejection of the existence of a physical effect. If the computed $\chi^2(\lg)$ becomes then greater than its critical value for a probability $p$, one has to check if residuals are smaller or bigger than those obtained by the reference solution. In the first case (smaller -- or better -- residuals), it means that the added effect increases significantly the quality of the residuals and is probably (with a probability $p$) a true physical effect. On the contrary, in the second case (bigger -- or degraded -- residuals), it means that the added effect is probably physically false. In our work, the critical increasing of $\chi^2(\lg)$ corresponds to a degradation of the residuals (see Supplemental Material for a detailed analysis). The massive graviton can then be rejected for high enough values of the mass (or low enough values of $\lg$). 

\section{Results}

\begin{figure}
        \includegraphics[scale=0.5]{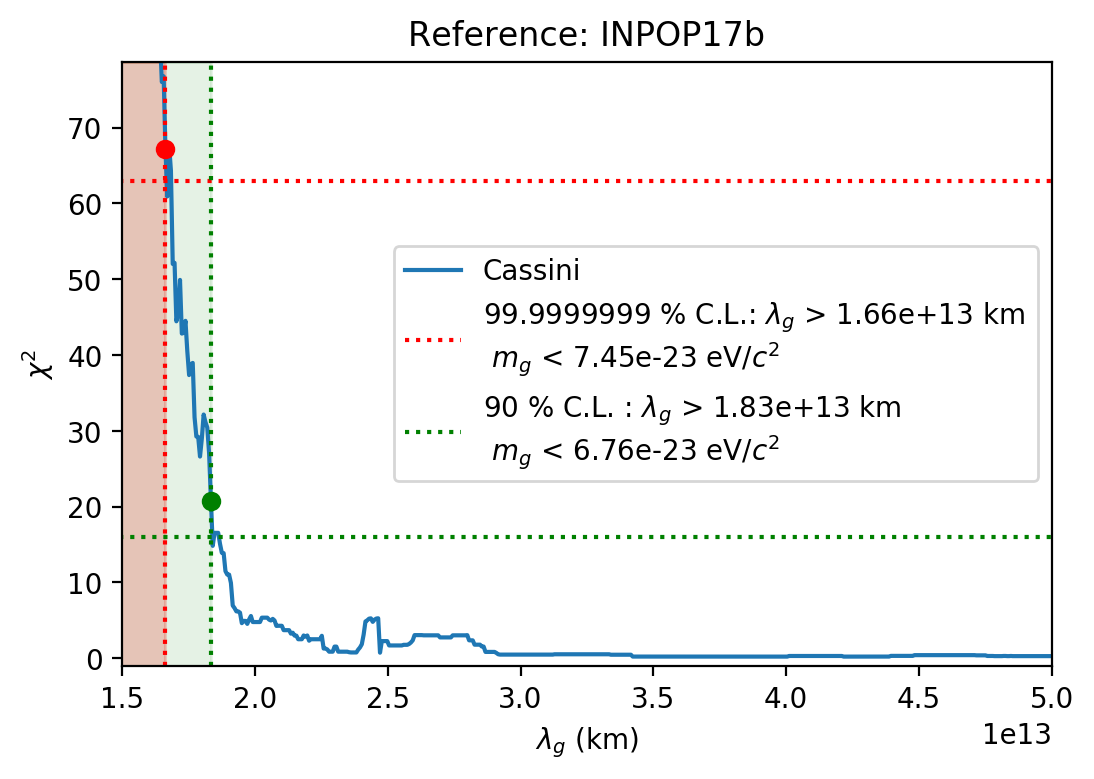}
        \includegraphics[scale=0.5]{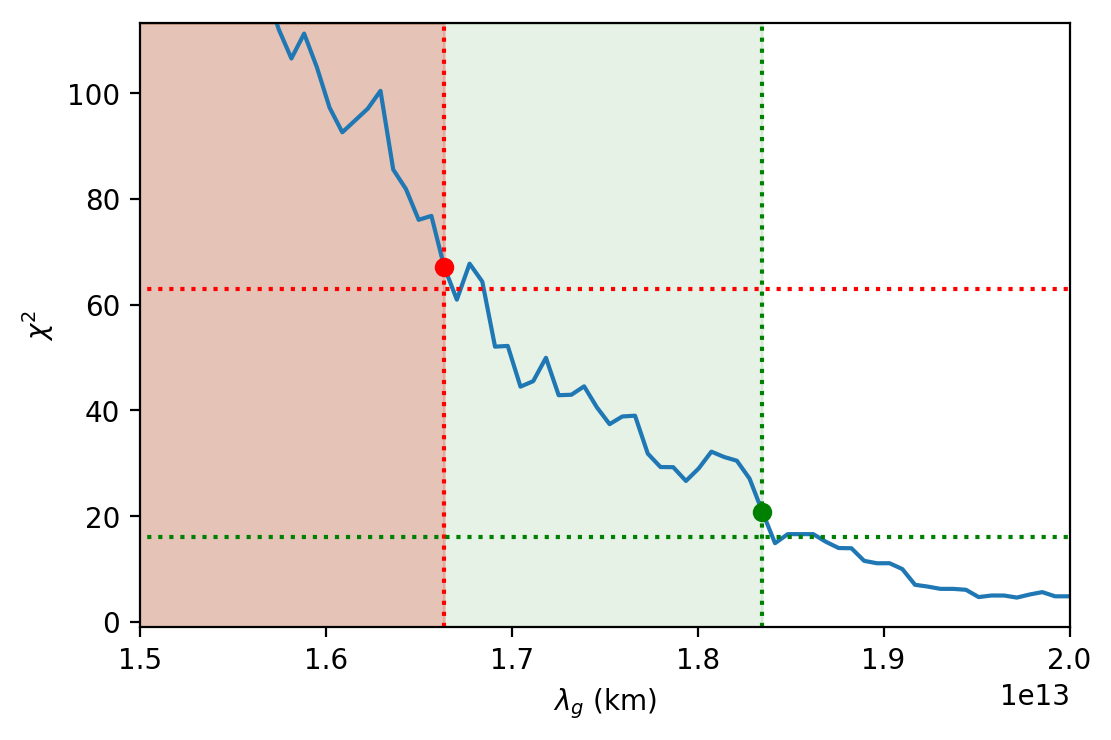}
        \caption{Plot of $\chi^2(\lg)$ and the constraints deduced for $\lg$. The probabilities $p=90\%$ and $p=99,9999999\%$ correspond to critical values of $\chi^2$ equal to respectively $15.99$ and $62.94$.}
        \label{fig:chi2}
\end{figure}

In Fig. \ref{fig:chi2} we plot the $\chi^2$ as a function of $\lg$. In this plot, we give two values of quantiles associated to two probabilities of significance, $p=90\%$ and $p=99,9999999\%$, which correspond to critical values of $\chi^2$ equal to $15.99$ and $62.94$ respectively for a 10 degrees of freedom $\chi^2$ distribution. We obtain respectively \colp{$\lambda_g > 1.83 \times 10^{13}$} km (resp. \colp{$m_g < 6.76 \times 10^{-23}$ eV/$c^2$}) and \colp{$\lambda_g > 1.66 \times 10^{13}$} km (resp. \colp{$m_g < 7.45 \times 10^{-23}$ eV/$c^2$}). These results are shown in Fig. \ref{fig:chi2}. We also provide a zoom of the main figure in order to show that the $\chi^2$ is not monotonic for small differences of $\lg$. However, if a given limit is crossed several times, our algorithm automatically takes the most conservative value in the discrete set of $\lg$, as can be seen in Fig. \ref{fig:chi2}.

---------------------------------------------------------------------
\section{Conclusion}

In the present manuscript, we deliver the first conservative estimate of the graviton mass from an actual fit of a combination of Solar System data, using a criterion based on a state of the art Solar System ephemerides: INPOP17b. The bound reads \colp{$\lambda_g > 1.83 \times 10^{13}$\,km (resp. $m_g < 6.76 \times 10^{-23}$ eV/$c^2$) with a confidence of 90\% and $\lambda_g > 1.66 \times 10^{13}$\,km (resp. $m_g < 7.45 \times 10^{-23}$ eV/$c^2$) with a confidence of 99.9999999\%.} 
As previously explained, in terms of a fifth force, the constraint on $\lambda_g$ can be translated into a constraint on $\lambda/\sqrt{\alpha}$, simply by substituting $\lambda_g$ by  $\lambda/\sqrt{\alpha}$, if $\alpha>0$.

The fact that our 90\% C.L. bound is comparable in magnitude to the one obtained by the LIGO-Virgo collaboration in the radiative regime \cite{ligo:2019ar,gw170104} is a pure coincidence: the two bounds rely on totally different types of observation --- gravitational waves versus radioscience in the Solar System --- and probe different aspects of the massive graviton phenomenology --- radiative versus Keplerian.


\begin{thebibliography}{21}%
\makeatletter
\providecommand \@ifxundefined [1]{%
 \@ifx{#1\undefined}
}%
\providecommand \@ifnum [1]{%
 \ifnum #1\expandafter \@firstoftwo
 \else \expandafter \@secondoftwo
 \fi
}%
\providecommand \@ifx [1]{%
 \ifx #1\expandafter \@firstoftwo
 \else \expandafter \@secondoftwo
 \fi
}%
\providecommand \natexlab [1]{#1}%
\providecommand \enquote  [1]{``#1''}%
\providecommand \bibnamefont  [1]{#1}%
\providecommand \bibfnamefont [1]{#1}%
\providecommand \citenamefont [1]{#1}%
\providecommand \href@noop [0]{\@secondoftwo}%
\providecommand \href [0]{\begingroup \@sanitize@url \@href}%
\providecommand \@href[1]{\@@startlink{#1}\@@href}%
\providecommand \@@href[1]{\endgroup#1\@@endlink}%
\providecommand \@sanitize@url [0]{\catcode `\\12\catcode `\$12\catcode
  `\&12\catcode `\#12\catcode `\^12\catcode `\_12\catcode `\%12\relax}%
\providecommand \@@startlink[1]{}%
\providecommand \@@endlink[0]{}%
\providecommand \url  [0]{\begingroup\@sanitize@url \@url }%
\providecommand \@url [1]{\endgroup\@href {#1}{\urlprefix }}%
\providecommand \urlprefix  [0]{URL }%
\providecommand \Eprint [0]{\href }%
\providecommand \doibase [0]{http://dx.doi.org/}%
\providecommand \selectlanguage [0]{\@gobble}%
\providecommand \bibinfo  [0]{\@secondoftwo}%
\providecommand \bibfield  [0]{\@secondoftwo}%
\providecommand \translation [1]{[#1]}%
\providecommand \BibitemOpen [0]{}%
\providecommand \bibitemStop [0]{}%
\providecommand \bibitemNoStop [0]{.\EOS\space}%
\providecommand \EOS [0]{\spacefactor3000\relax}%
\providecommand \BibitemShut  [1]{\csname bibitem#1\endcsname}%
\let\auto@bib@innerbib\@empty
\bibitem [{\citenamefont {{Fierz}}\ and\ \citenamefont
  {{Pauli}}(1939)}]{fierz:1939rs}%
  \BibitemOpen
  \bibfield  {author} {\bibinfo {author} {\bibfnamefont {M.}~\bibnamefont
  {{Fierz}}}\ and\ \bibinfo {author} {\bibfnamefont {W.}~\bibnamefont
  {{Pauli}}},\ }\bibfield  {title} {\enquote {\bibinfo {title} {{On
  Relativistic Wave Equations for Particles of Arbitrary Spin in an
  Electromagnetic Field}},}\ }\href {\doibase 10.1098/rspa.1939.0140}
  {\bibfield  {journal} {\bibinfo  {journal} {Proceedings of the Royal Society
  of London Series A}\ }\textbf {\bibinfo {volume} {173}},\ \bibinfo {pages}
  {211--232} (\bibinfo {year} {1939})}\BibitemShut {NoStop}%
\bibitem [{\citenamefont {{de Rham}}\ \emph {et~al.}(2017)\citenamefont {{de
  Rham}}, \citenamefont {{Deskins}}, \citenamefont {{Tolley}},\ and\
  \citenamefont {{Zhou}}}]{derham:2017rp}%
  \BibitemOpen
  \bibfield  {author} {\bibinfo {author} {\bibfnamefont {C.}~\bibnamefont {{de
  Rham}}}, \bibinfo {author} {\bibfnamefont {J.~T.}\ \bibnamefont {{Deskins}}},
  \bibinfo {author} {\bibfnamefont {A.~J.}\ \bibnamefont {{Tolley}}}, \ and\
  \bibinfo {author} {\bibfnamefont {S.-Y.}\ \bibnamefont {{Zhou}}},\ }\bibfield
   {title} {\enquote {\bibinfo {title} {{Graviton mass bounds}},}\ }\href
  {\doibase 10.1103/RevModPhys.89.025004} {\bibfield  {journal} {\bibinfo
  {journal} {Reviews of Modern Physics}\ }\textbf {\bibinfo {volume} {89}},\
  \bibinfo {eid} {025004} (\bibinfo {year} {2017})},\ \Eprint
  {http://arxiv.org/abs/1606.08462} {arXiv:1606.08462} \BibitemShut {NoStop}%
\bibitem [{\citenamefont {{de Rham}}(2014)}]{derham:2014lr}%
  \BibitemOpen
  \bibfield  {author} {\bibinfo {author} {\bibfnamefont {C.}~\bibnamefont {{de
  Rham}}},\ }\bibfield  {title} {\enquote {\bibinfo {title} {{Massive
  Gravity}},}\ }\href {\doibase 10.12942/lrr-2014-7} {\bibfield  {journal}
  {\bibinfo  {journal} {Living Reviews in Relativity}\ }\textbf {\bibinfo
  {volume} {17}},\ \bibinfo {eid} {7} (\bibinfo {year} {2014})},\ \Eprint
  {http://arxiv.org/abs/1401.4173} {arXiv:1401.4173 [hep-th]} \BibitemShut
  {NoStop}%
\bibitem [{\citenamefont {{The LIGO Scientific Collaboration}}\ \emph
  {et~al.}(2018)\citenamefont {{The LIGO Scientific Collaboration}},
  \citenamefont {{the Virgo Collaboration}}, \citenamefont {{Abbott}},
  \citenamefont {{Abbott}}, \citenamefont {{Abbott}}, \citenamefont
  {{Abraham}}, \citenamefont {{Acernese}}, \citenamefont {{Ackley}},
  \citenamefont {{Adams}}, \citenamefont {{Adhikari}},\ and\ \citenamefont
  {et~al.}}]{ligo:2018ax}%
  \BibitemOpen
  \bibfield  {author} {\bibinfo {author} {\bibnamefont {{The LIGO Scientific
  Collaboration}}}, \bibinfo {author} {\bibnamefont {{the Virgo
  Collaboration}}}, \bibinfo {author} {\bibfnamefont {B.~P.}\ \bibnamefont
  {{Abbott}}}, \bibinfo {author} {\bibfnamefont {R.}~\bibnamefont {{Abbott}}},
  \bibinfo {author} {\bibfnamefont {T.~D.}\ \bibnamefont {{Abbott}}}, \bibinfo
  {author} {\bibfnamefont {S.}~\bibnamefont {{Abraham}}}, \bibinfo {author}
  {\bibfnamefont {F.}~\bibnamefont {{Acernese}}}, \bibinfo {author}
  {\bibfnamefont {K.}~\bibnamefont {{Ackley}}}, \bibinfo {author}
  {\bibfnamefont {C.}~\bibnamefont {{Adams}}}, \bibinfo {author} {\bibfnamefont
  {R.~X.}\ \bibnamefont {{Adhikari}}}, \ and\ \bibinfo {author} {\bibnamefont
  {et~al.}},\ }\bibfield  {title} {\enquote {\bibinfo {title} {{GWTC-1: A
  Gravitational-Wave Transient Catalog of Compact Binary Mergers Observed by
  LIGO and Virgo during the First and Second Observing Runs}},}\ }\href@noop {}
  {\bibfield  {journal} {\bibinfo  {journal} {arXiv e-prints}\ } (\bibinfo
  {year} {2018})},\ \Eprint {http://arxiv.org/abs/1811.12907} {arXiv:1811.12907
  [astro-ph.HE]} \BibitemShut {NoStop}%
\bibitem [{\citenamefont {{Will}}(1998)}]{will:1998pr}%
  \BibitemOpen
  \bibfield  {author} {\bibinfo {author} {\bibfnamefont {C.~M.}\ \bibnamefont
  {{Will}}},\ }\bibfield  {title} {\enquote {\bibinfo {title} {{Bounding the
  mass of the graviton using gravitational-wave observations of inspiralling
  compact binaries}},}\ }\href {\doibase 10.1103/PhysRevD.57.2061} {\bibfield
  {journal} {\bibinfo  {journal} {\prd}\ }\textbf {\bibinfo {volume} {57}},\
  \bibinfo {pages} {2061--2068} (\bibinfo {year} {1998})},\ \Eprint
  {http://arxiv.org/abs/gr-qc/9709011} {gr-qc/9709011} \BibitemShut {NoStop}%
\bibitem [{\citenamefont {{Del Pozzo}}\ \emph {et~al.}(2011)\citenamefont {{Del
  Pozzo}}, \citenamefont {{Veitch}},\ and\ \citenamefont
  {{Vecchio}}}]{delpozzo:2011pr}%
  \BibitemOpen
  \bibfield  {author} {\bibinfo {author} {\bibfnamefont {W.}~\bibnamefont {{Del
  Pozzo}}}, \bibinfo {author} {\bibfnamefont {J.}~\bibnamefont {{Veitch}}}, \
  and\ \bibinfo {author} {\bibfnamefont {A.}~\bibnamefont {{Vecchio}}},\
  }\bibfield  {title} {\enquote {\bibinfo {title} {{Testing general relativity
  using Bayesian model selection: Applications to observations of gravitational
  waves from compact binary systems}},}\ }\href {\doibase
  10.1103/PhysRevD.83.082002} {\bibfield  {journal} {\bibinfo  {journal}
  {\prd}\ }\textbf {\bibinfo {volume} {83}},\ \bibinfo {eid} {082002} (\bibinfo
  {year} {2011})},\ \Eprint {http://arxiv.org/abs/1101.1391} {arXiv:1101.1391
  [gr-qc]} \BibitemShut {NoStop}%
\bibitem [{\citenamefont {{The LIGO Scientific Collaboration}}\ and\
  \citenamefont {{the Virgo Collaboration}}(2019)}]{ligo:2019ar}%
  \BibitemOpen
  \bibfield  {author} {\bibinfo {author} {\bibnamefont {{The LIGO Scientific
  Collaboration}}}\ and\ \bibinfo {author} {\bibnamefont {{the Virgo
  Collaboration}}},\ }\bibfield  {title} {\enquote {\bibinfo {title} {{Tests of
  General Relativity with the Binary Black Hole Signals from the LIGO-Virgo
  Catalog GWTC-1}},}\ }\href@noop {} {\bibfield  {journal} {\bibinfo  {journal}
  {arXiv e-prints}\ ,\ \bibinfo {eid} {arXiv:1903.04467}} (\bibinfo {year}
  {2019})},\ \Eprint {http://arxiv.org/abs/1903.04467} {arXiv:1903.04467
  [gr-qc]} \BibitemShut {NoStop}%
\bibitem [{\citenamefont {Will}(2018)}]{will:2018cq}%
  \BibitemOpen
  \bibfield  {author} {\bibinfo {author} {\bibfnamefont {Clifford~M}\
  \bibnamefont {Will}},\ }\bibfield  {title} {\enquote {\bibinfo {title} {Solar
  system versus gravitational-wave bounds on the graviton mass},}\ }\href
  {http://stacks.iop.org/0264-9381/35/i=17/a=17LT01} {\bibfield  {journal}
  {\bibinfo  {journal} {Classical and Quantum Gravity}\ }\textbf {\bibinfo
  {volume} {35}},\ \bibinfo {pages} {17LT01} (\bibinfo {year}
  {2018})}\BibitemShut {NoStop}%
\bibitem [{\citenamefont {{Fienga}}\ \emph {et~al.}(2008)\citenamefont
  {{Fienga}}, \citenamefont {{Manche}}, \citenamefont {{Laskar}},\ and\
  \citenamefont {{Gastineau}}}]{fienga:2008aa}%
  \BibitemOpen
  \bibfield  {author} {\bibinfo {author} {\bibfnamefont {A.}~\bibnamefont
  {{Fienga}}}, \bibinfo {author} {\bibfnamefont {H.}~\bibnamefont {{Manche}}},
  \bibinfo {author} {\bibfnamefont {J.}~\bibnamefont {{Laskar}}}, \ and\
  \bibinfo {author} {\bibfnamefont {M.}~\bibnamefont {{Gastineau}}},\
  }\bibfield  {title} {\enquote {\bibinfo {title} {{INPOP06: a new numerical
  planetary ephemeris}},}\ }\href {\doibase 10.1051/0004-6361:20066607}
  {\bibfield  {journal} {\bibinfo  {journal} {\aap}\ }\textbf {\bibinfo
  {volume} {477}},\ \bibinfo {pages} {315--327} (\bibinfo {year}
  {2008})}\BibitemShut {NoStop}%
\bibitem [{\citenamefont {Moyer}(2003)}]{moyer:2003}%
  \BibitemOpen
  \bibfield  {author} {\bibinfo {author} {\bibfnamefont {T.~D.}\ \bibnamefont
  {Moyer}},\ }\href {\doibase 10.1002/0471728470} {\emph {\bibinfo {title}
  {Deep Space Communications and Navigation Series}}},\ Vol.~\bibinfo {volume}
  {2}\ (\bibinfo  {publisher} {John Wiley {\&} Sons, Inc.},\ \bibinfo {address}
  {Hoboken, NJ, USA},\ \bibinfo {year} {2003})\BibitemShut {NoStop}%
\bibitem [{\citenamefont {{Fienga}}\ \emph {et~al.}(2011)\citenamefont
  {{Fienga}}, \citenamefont {{Laskar}}, \citenamefont {{Kuchynka}},
  \citenamefont {{Manche}}, \citenamefont {{Desvignes}}, \citenamefont
  {{Gastineau}}, \citenamefont {{Cognard}},\ and\ \citenamefont
  {{Theureau}}}]{fienga:2011cm}%
  \BibitemOpen
  \bibfield  {author} {\bibinfo {author} {\bibfnamefont {A.}~\bibnamefont
  {{Fienga}}}, \bibinfo {author} {\bibfnamefont {J.}~\bibnamefont {{Laskar}}},
  \bibinfo {author} {\bibfnamefont {P.}~\bibnamefont {{Kuchynka}}}, \bibinfo
  {author} {\bibfnamefont {H.}~\bibnamefont {{Manche}}}, \bibinfo {author}
  {\bibfnamefont {G.}~\bibnamefont {{Desvignes}}}, \bibinfo {author}
  {\bibfnamefont {M.}~\bibnamefont {{Gastineau}}}, \bibinfo {author}
  {\bibfnamefont {I.}~\bibnamefont {{Cognard}}}, \ and\ \bibinfo {author}
  {\bibfnamefont {G.}~\bibnamefont {{Theureau}}},\ }\bibfield  {title}
  {\enquote {\bibinfo {title} {{The INPOP10a planetary ephemeris and its
  applications in fundamental physics}},}\ }\href {\doibase
  10.1007/s10569-011-9377-8} {\bibfield  {journal} {\bibinfo  {journal}
  {Celestial Mechanics and Dynamical Astronomy}\ }\textbf {\bibinfo {volume}
  {111}},\ \bibinfo {pages} {363--385} (\bibinfo {year} {2011})},\ \Eprint
  {http://arxiv.org/abs/1108.5546} {arXiv:1108.5546 [astro-ph.EP]} \BibitemShut
  {NoStop}%
\bibitem [{\citenamefont {{Verma}}\ \emph {et~al.}(2014)\citenamefont
  {{Verma}}, \citenamefont {{Fienga}}, \citenamefont {{Laskar}}, \citenamefont
  {{Manche}},\ and\ \citenamefont {{Gastineau}}}]{verma:2014aa}%
  \BibitemOpen
  \bibfield  {author} {\bibinfo {author} {\bibfnamefont {A.~K.}\ \bibnamefont
  {{Verma}}}, \bibinfo {author} {\bibfnamefont {A.}~\bibnamefont {{Fienga}}},
  \bibinfo {author} {\bibfnamefont {J.}~\bibnamefont {{Laskar}}}, \bibinfo
  {author} {\bibfnamefont {H.}~\bibnamefont {{Manche}}}, \ and\ \bibinfo
  {author} {\bibfnamefont {M.}~\bibnamefont {{Gastineau}}},\ }\bibfield
  {title} {\enquote {\bibinfo {title} {{Use of MESSENGER radioscience data to
  improve planetary ephemeris and to test general relativity}},}\ }\href
  {\doibase 10.1051/0004-6361/201322124} {\bibfield  {journal} {\bibinfo
  {journal} {\aap}\ }\textbf {\bibinfo {volume} {561}},\ \bibinfo {eid} {A115}
  (\bibinfo {year} {2014})},\ \Eprint {http://arxiv.org/abs/1306.5569}
  {arXiv:1306.5569 [astro-ph.EP]} \BibitemShut {NoStop}%
\bibitem [{\citenamefont {{Fienga}}\ \emph {et~al.}(2015)\citenamefont
  {{Fienga}}, \citenamefont {{Laskar}}, \citenamefont {{Exertier}},
  \citenamefont {{Manche}},\ and\ \citenamefont {{Gastineau}}}]{fienga:2015cm}%
  \BibitemOpen
  \bibfield  {author} {\bibinfo {author} {\bibfnamefont {A.}~\bibnamefont
  {{Fienga}}}, \bibinfo {author} {\bibfnamefont {J.}~\bibnamefont {{Laskar}}},
  \bibinfo {author} {\bibfnamefont {P.}~\bibnamefont {{Exertier}}}, \bibinfo
  {author} {\bibfnamefont {H.}~\bibnamefont {{Manche}}}, \ and\ \bibinfo
  {author} {\bibfnamefont {M.}~\bibnamefont {{Gastineau}}},\ }\bibfield
  {title} {\enquote {\bibinfo {title} {{Numerical estimation of the sensitivity
  of INPOP planetary ephemerides to general relativity parameters}},}\ }\href
  {\doibase 10.1007/s10569-015-9639-y} {\bibfield  {journal} {\bibinfo
  {journal} {Celestial Mechanics and Dynamical Astronomy}\ }\textbf {\bibinfo
  {volume} {123}},\ \bibinfo {pages} {325--349} (\bibinfo {year}
  {2015})}\BibitemShut {NoStop}%
\bibitem [{\citenamefont {{Viswanathan}}\ \emph {et~al.}(2018)\citenamefont
  {{Viswanathan}}, \citenamefont {{Fienga}}, \citenamefont {{Minazzoli}},
  \citenamefont {{Bernus}}, \citenamefont {{Laskar}},\ and\ \citenamefont
  {{Gastineau}}}]{viswanathan:2018mn}%
  \BibitemOpen
  \bibfield  {author} {\bibinfo {author} {\bibfnamefont {V.}~\bibnamefont
  {{Viswanathan}}}, \bibinfo {author} {\bibfnamefont {A.}~\bibnamefont
  {{Fienga}}}, \bibinfo {author} {\bibfnamefont {O.}~\bibnamefont
  {{Minazzoli}}}, \bibinfo {author} {\bibfnamefont {L.}~\bibnamefont
  {{Bernus}}}, \bibinfo {author} {\bibfnamefont {J.}~\bibnamefont {{Laskar}}},
  \ and\ \bibinfo {author} {\bibfnamefont {M.}~\bibnamefont {{Gastineau}}},\
  }\bibfield  {title} {\enquote {\bibinfo {title} {{The new lunar ephemeris
  INPOP17a and its application to fundamental physics}},}\ }\href {\doibase
  10.1093/mnras/sty096} {\bibfield  {journal} {\bibinfo  {journal} {\mnras}\
  }\textbf {\bibinfo {volume} {476}},\ \bibinfo {pages} {1877--1888} (\bibinfo
  {year} {2018})},\ \Eprint {http://arxiv.org/abs/1710.09167} {arXiv:1710.09167
  [gr-qc]} \BibitemShut {NoStop}%
\bibitem [{\citenamefont {{Viswanathan}}\ \emph {et~al.}(2017)\citenamefont
  {{Viswanathan}}, \citenamefont {{Fienga}}, \citenamefont {{Gastineau}},\ and\
  \citenamefont {{Laskar}}}]{viswanathan:2018dc}%
  \BibitemOpen
  \bibfield  {author} {\bibinfo {author} {\bibfnamefont {V.}~\bibnamefont
  {{Viswanathan}}}, \bibinfo {author} {\bibfnamefont {A.}~\bibnamefont
  {{Fienga}}}, \bibinfo {author} {\bibfnamefont {M.}~\bibnamefont
  {{Gastineau}}}, \ and\ \bibinfo {author} {\bibfnamefont {J.}~\bibnamefont
  {{Laskar}}},\ }\bibfield  {title} {\enquote {\bibinfo {title} {{INPOP17a
  planetary ephemerides}},}\ }\href@noop {} {\bibfield  {journal} {\bibinfo
  {journal} {Notes Scientifiques et Techniques de l'Institut de Mecanique
  Celeste}\ }\textbf {\bibinfo {volume} {108}} (\bibinfo {year} {2017})},\
  \bibinfo {note} {last Accessed: 2018-11-13}\BibitemShut {NoStop}%
\bibitem [{\citenamefont {{Verma}}\ and\ \citenamefont
  {{Margot}}(2016)}]{2016JGRE..121.1627V}%
  \BibitemOpen
  \bibfield  {author} {\bibinfo {author} {\bibfnamefont {A.~K.}\ \bibnamefont
  {{Verma}}}\ and\ \bibinfo {author} {\bibfnamefont {J.-L.}\ \bibnamefont
  {{Margot}}},\ }\bibfield  {title} {\enquote {\bibinfo {title} {{Mercury's
  gravity, tides, and spin from MESSENGER radio science data}},}\ }\href
  {\doibase 10.1002/2016JE005037} {\bibfield  {journal} {\bibinfo  {journal}
  {Journal of Geophysical Research (Planets)}\ }\textbf {\bibinfo {volume}
  {121}},\ \bibinfo {pages} {1627--1640} (\bibinfo {year} {2016})},\ \Eprint
  {http://arxiv.org/abs/1608.01360} {arXiv:1608.01360 [astro-ph.EP]}
  \BibitemShut {NoStop}%
\bibitem [{\citenamefont {{Talmadge}}\ \emph {et~al.}(1988)\citenamefont
  {{Talmadge}}, \citenamefont {{Berthias}}, \citenamefont {{Hellings}},\ and\
  \citenamefont {{Standish}}}]{talmadge:1988pl}%
  \BibitemOpen
  \bibfield  {author} {\bibinfo {author} {\bibfnamefont {C.}~\bibnamefont
  {{Talmadge}}}, \bibinfo {author} {\bibfnamefont {J.-P.}\ \bibnamefont
  {{Berthias}}}, \bibinfo {author} {\bibfnamefont {R.~W.}\ \bibnamefont
  {{Hellings}}}, \ and\ \bibinfo {author} {\bibfnamefont {E.~M.}\ \bibnamefont
  {{Standish}}},\ }\bibfield  {title} {\enquote {\bibinfo {title}
  {{Model-independent constraints on possible modifications of Newtonian
  gravity}},}\ }\href {\doibase 10.1103/PhysRevLett.61.1159} {\bibfield
  {journal} {\bibinfo  {journal} {Physical Review Letters}\ }\textbf {\bibinfo
  {volume} {61}},\ \bibinfo {pages} {1159--1162} (\bibinfo {year}
  {1988})}\BibitemShut {NoStop}%
\bibitem [{\citenamefont {{Hees}}\ \emph {et~al.}(2017)\citenamefont {{Hees}},
  \citenamefont {{Do}}, \citenamefont {{Ghez}}, \citenamefont {{Martinez}},
  \citenamefont {{Naoz}}, \citenamefont {{Becklin}}, \citenamefont {{Boehle}},
  \citenamefont {{Chappell}}, \citenamefont {{Chu}}, \citenamefont
  {{Dehghanfar}}, \citenamefont {{Kosmo}}, \citenamefont {{Lu}}, \citenamefont
  {{Matthews}}, \citenamefont {{Morris}}, \citenamefont {{Sakai}},
  \citenamefont {{Sch{\"o}del}},\ and\ \citenamefont
  {{Witzel}}}]{hees:2017prl}%
  \BibitemOpen
  \bibfield  {author} {\bibinfo {author} {\bibfnamefont {A.}~\bibnamefont
  {{Hees}}}, \bibinfo {author} {\bibfnamefont {T.}~\bibnamefont {{Do}}},
  \bibinfo {author} {\bibfnamefont {A.~M.}\ \bibnamefont {{Ghez}}}, \bibinfo
  {author} {\bibfnamefont {G.~D.}\ \bibnamefont {{Martinez}}}, \bibinfo
  {author} {\bibfnamefont {S.}~\bibnamefont {{Naoz}}}, \bibinfo {author}
  {\bibfnamefont {E.~E.}\ \bibnamefont {{Becklin}}}, \bibinfo {author}
  {\bibfnamefont {A.}~\bibnamefont {{Boehle}}}, \bibinfo {author}
  {\bibfnamefont {S.}~\bibnamefont {{Chappell}}}, \bibinfo {author}
  {\bibfnamefont {D.}~\bibnamefont {{Chu}}}, \bibinfo {author} {\bibfnamefont
  {A.}~\bibnamefont {{Dehghanfar}}}, \bibinfo {author} {\bibfnamefont
  {K.}~\bibnamefont {{Kosmo}}}, \bibinfo {author} {\bibfnamefont {J.~R.}\
  \bibnamefont {{Lu}}}, \bibinfo {author} {\bibfnamefont {K.}~\bibnamefont
  {{Matthews}}}, \bibinfo {author} {\bibfnamefont {M.~R.}\ \bibnamefont
  {{Morris}}}, \bibinfo {author} {\bibfnamefont {S.}~\bibnamefont {{Sakai}}},
  \bibinfo {author} {\bibfnamefont {R.}~\bibnamefont {{Sch{\"o}del}}}, \ and\
  \bibinfo {author} {\bibfnamefont {G.}~\bibnamefont {{Witzel}}},\ }\bibfield
  {title} {\enquote {\bibinfo {title} {{Testing General Relativity with Stellar
  Orbits around the Supermassive Black Hole in Our Galactic Center}},}\ }\href
  {\doibase 10.1103/PhysRevLett.118.211101} {\bibfield  {journal} {\bibinfo
  {journal} {Physical Review Letters}\ }\textbf {\bibinfo {volume} {118}},\
  \bibinfo {eid} {211101} (\bibinfo {year} {2017})},\ \Eprint
  {http://arxiv.org/abs/1705.07902} {arXiv:1705.07902} \BibitemShut {NoStop}%
\bibitem [{\citenamefont {Pearson}(1992)}]{Pearson:1992}%
  \BibitemOpen
  \bibfield  {author} {\bibinfo {author} {\bibfnamefont {Karl}\ \bibnamefont
  {Pearson}},\ }\enquote {\bibinfo {title} {On the criterion that a given
  system of deviations from the probable in the case of a correlated system of
  variables is such that it can be reasonably supposed to have arisen from
  random sampling},}\ in\ \href {\doibase 10.1007/978-1-4612-4380-9_2} {\emph
  {\bibinfo {booktitle} {Breakthroughs in Statistics: Methodology and
  Distribution}}},\ \bibinfo {editor} {edited by\ \bibinfo {editor}
  {\bibfnamefont {Samuel}\ \bibnamefont {Kotz}}\ and\ \bibinfo {editor}
  {\bibfnamefont {Norman~L.}\ \bibnamefont {Johnson}}}\ (\bibinfo  {publisher}
  {Springer New York},\ \bibinfo {address} {New York, NY},\ \bibinfo {year}
  {1992})\ pp.\ \bibinfo {pages} {11--28}\BibitemShut {NoStop}%
\bibitem [{\citenamefont {SCOTT}(1979)}]{scott:1979bm}%
  \BibitemOpen
  \bibfield  {author} {\bibinfo {author} {\bibfnamefont {DAVID~W.}\
  \bibnamefont {SCOTT}},\ }\bibfield  {title} {\enquote {\bibinfo {title} {On
  optimal and data-based histograms},}\ }\href {\doibase
  10.1093/biomet/66.3.605} {\bibfield  {journal} {\bibinfo  {journal}
  {Biometrika}\ }\textbf {\bibinfo {volume} {66}},\ \bibinfo {pages} {605--610}
  (\bibinfo {year} {1979})}\BibitemShut {NoStop}%
\bibitem [{\citenamefont {{Abbott}}\ \emph {et~al.}(2017)\citenamefont
  {{Abbott}}, \citenamefont {{Abbott}}, \citenamefont {{Abbott}}, \citenamefont
  {{Acernese}}, \citenamefont {{Ackley}}, \citenamefont {{Adams}},
  \citenamefont {{Adams}}, \citenamefont {{Addesso}}, \citenamefont
  {{Adhikari}}, \citenamefont {{Adya}},\ and\ \citenamefont
  {et~al.}}]{gw170104}%
  \BibitemOpen
  \bibfield  {author} {\bibinfo {author} {\bibfnamefont {B.~P.}\ \bibnamefont
  {{Abbott}}}, \bibinfo {author} {\bibfnamefont {R.}~\bibnamefont {{Abbott}}},
  \bibinfo {author} {\bibfnamefont {T.~D.}\ \bibnamefont {{Abbott}}}, \bibinfo
  {author} {\bibfnamefont {F.}~\bibnamefont {{Acernese}}}, \bibinfo {author}
  {\bibfnamefont {K.}~\bibnamefont {{Ackley}}}, \bibinfo {author}
  {\bibfnamefont {C.}~\bibnamefont {{Adams}}}, \bibinfo {author} {\bibfnamefont
  {T.}~\bibnamefont {{Adams}}}, \bibinfo {author} {\bibfnamefont
  {P.}~\bibnamefont {{Addesso}}}, \bibinfo {author} {\bibfnamefont {R.~X.}\
  \bibnamefont {{Adhikari}}}, \bibinfo {author} {\bibfnamefont {V.~B.}\
  \bibnamefont {{Adya}}}, \ and\ \bibinfo {author} {\bibnamefont {et~al.}},\
  }\bibfield  {title} {\enquote {\bibinfo {title} {{GW170104: Observation of a
  50-Solar-Mass Binary Black Hole Coalescence at Redshift 0.2}},}\ }\href
  {\doibase 10.1103/PhysRevLett.118.221101} {\bibfield  {journal} {\bibinfo
  {journal} {Physical Review Letters}\ }\textbf {\bibinfo {volume} {118}},\
  \bibinfo {eid} {221101} (\bibinfo {year} {2017})},\ \Eprint
  {http://arxiv.org/abs/1706.01812} {arXiv:1706.01812 [gr-qc]} \BibitemShut
  {NoStop}%
\end{thebibliography}
\end{document}


\title{Supplemental materials: Constraining the mass of the graviton with the planetary ephemeris INPOP}

\author{
L. Bernus $^{1}$,
O. Minazzoli $^{3,4}$,
A. Fienga $^{2,1}$,
M. Gastineau $^{1}$,
J. Laskar $^{1}$,
P. Deram $^2$\\
$^{1}$IMCCE, Observatoire de Paris, PSL University, CNRS, Sorbonne Universit\'e, 77 avenue Denfert-Rochereau, 75014 Paris, France \\
$^{2}$G\'eoazur, Observatoire de la C\^ote d'Azur, Universit\'e C\^ote d'Azur, IRD, 250 Rue Albert Einstein, 06560 Valbonne, France\\
$^{3}$Centre Scientifique de Monaco, 8 Quai Antoine 1er, Monaco \\
$^{4}$Artemis, Universit\'e C\^ote d'Azur, CNRS, Observatoire de la C\^ote d'Azur, BP4229, 06304, Nice Cedex 4, France
}


\maketitle

\section{Numerical analysis}\label{sec:num}

To model and confront the massive graviton to Solar System observations, we add its contribution 
to the INPOP17b Solar System model (including asteroid masses), and then fit the newly obtained ephemeris according to the procedure described in \cite{viswanathan:2018dc}. For a fixed value of $\lg$, we fit the parameters of the whole model to the data.

\begin{figure}
  	\includegraphics[scale=0.5]{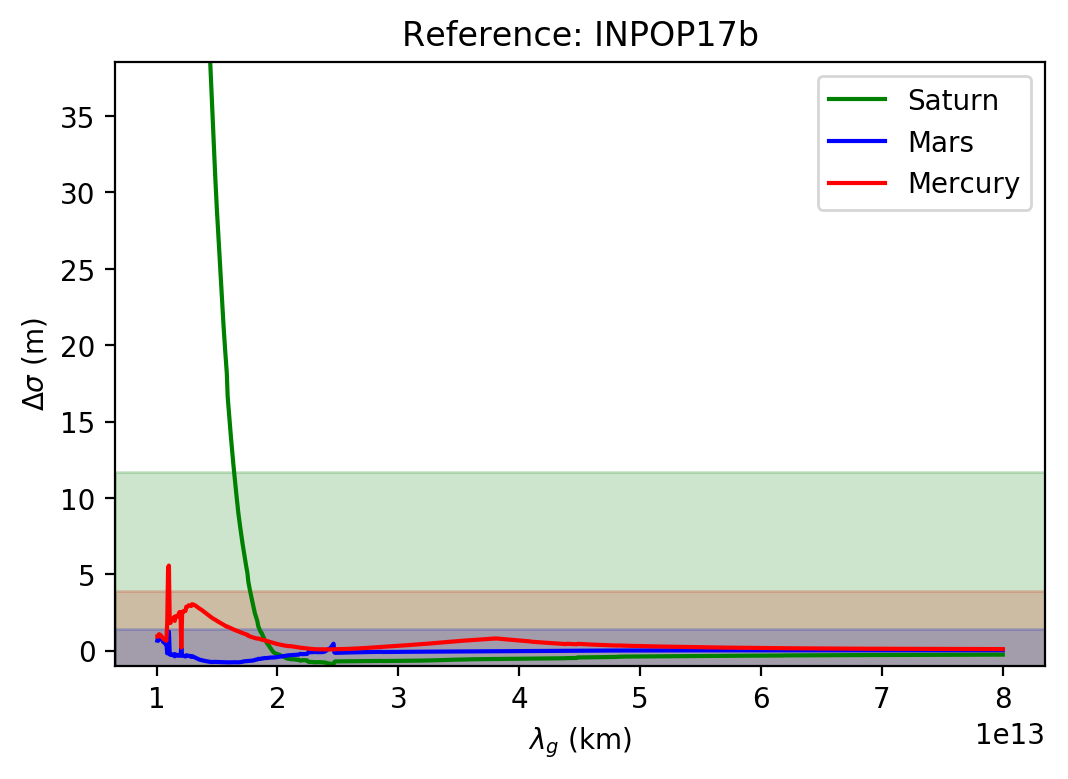}
  	\caption{Plot of the standard deviations with respect to $\lg$ after 12 iterations of the data fit. The values of the reference solution INPOP17b standard deviations (given in Table \ref{table:sigmavalues}) have been removed. The colored areas correspond to zones where the standard deviations of each observable are considered a priori as being marginally to not significant. The limits come from the estimation of the internal accuracy of the ephemeris obtained through the comparison with DE436, see Table \ref{table:sigmavalues} as well as \colp{text below}. [A rigorous evaluation of the significance of the deterioration is made in the main part of the manuscript.]}
  	\label{fig:sigma}
\end{figure}

We have performed an iterative fit of the INPOP17b parameters of the ephemeris for each given values of $\lg$. In order to compare the tested and the reference solutions, one needs to apply exactly the same procedure in each case. Accordingly, the same data and weights are used. \colp{As usual, weights are representative of data uncertainty and distribution}. The comparison between the residuals of a solution with a massive graviton to the residuals of the reference solution gives a measurement of the sensibility of the ephemeris to the effects of a massive graviton. For each value of $\lg$ we compute the standard deviation of the residuals of range observations. More specifically, we exhibit residuals obtained with observations for Cassini mission, Messenger mission, and Mars Odyssey and Mex mission. Other observations are less relevant due to less accurate data and/or high correlation with $\lambda_g$. This algorithm processes for \colp{1024} different fixed values of $\lg$ between 1$\times10^{13}$ and 8 $\times10^{13}$\,km. We plot the different standard deviations with respect to $\lg$ for each iteration. We remove the values of the reference solution standard deviations \footnote{i.e. for $\lambda_g =+\infty$} listed in Table \ref{table:sigmavalues}. In Table \ref{table:sigmavalues}, we indeed give the 1$\sigma$ standard deviations of INPOP17b residuals together with the 1$\sigma$ differences obtained between INPOP17b and the Jet Propulsion Laboratory ephemeris DE436 \footnote{Which is based on DE430 (\cite{folkner:2014in}).} geocentric distances on the time interval of the data sample. The latter gives an idea of the internal accuracy of the reference ephemeris itself. At about the $10$th iteration, the standard deviations for the three sets of residuals stop evolving --- meaning that the adjustment has converged. We report the plot of the standard deviations of the last iteration (the 13th) in Fig. \ref{fig:sigma}.

\begin{table*}

	\begin{tabular}{l|l c | c | c}
	\hline
		Observations & Time  & $\#$ & (O-C) INPOP17b & INPOP17b-DE436\\
		& Intervals & & 1$\sigma$ & 1$\sigma$ \\
		& & & [m] & [m]\\
		\hline
		Mercury (Messenger) & 2011 : 2013.2  & 950 & 7.2 & 3.9  \\
		Mars (Ody, MEX) & 2002 : 2016.4 & 52946 & 5.0 & 1.4 \\
		Saturn (Cassini) & 2004 : 2014 & 175 & 32.1 & 11.7 \\
		\hline
	\end{tabular}
	\caption{Summary of data selected for monitoring INPOP sensitivity to $\lg$. Columns 2 and 3 provide the time coverage of the sample and the number of observations per sample respectively. Column 4 gives the 1$\sigma$ standard deviation of residuals obtained with INPOP17b. These standard deviations are taken as reference values in the text. The last column indicates the 1$\sigma$ differences between INPOP17b and DE436 for geocentric distances and for interval of time covered by the two ephemeris adjustments.}
	\label{table:sigmavalues}
\end{table*}

An important point to have in mind is that Mars data constrain the global fit of parameters to observations --- thanks to the important weight in the fit given to the Mars Odyssey and MEX missions accurate data in the reference solution \footnote{As mentioned above, weights are representative of data uncertainty and distribution.}. So, while in Will's analysis \cite{will:2018cq}, the high quality of Martian data is expected to allow the best constraints on $\lambda_g$, this high quality actually helps to better adjust the whole set of parameters --- but they do not significantly constrain $\lambda_g$. Alternatively, given the fact that Saturn's semi-major axe is less correlated to $\lambda_g$ (see Table \colp{I} in the main part of the manuscript), it is not surprising to see in Fig. \ref{fig:sigma} that Saturn positions deduced from the Cassini observations are actually the most constraining on $\lambda_g$.

It is found that the variation of the standard deviations of the selected residuals dominates compared to the variations of the mean values of the residuals, for all values of $\lg$. Therefore, in the following, we focus on standard deviations of data.

As one can see in Fig. \ref{fig:sigma}, after 12 iterations, only residuals deduced from Saturn positions obtained with Cassini show important deviations at a high value of $\lg$. On the other hand, all the values of standard deviation of Mars data are below 1.5 m higher than the reference value. Around $\lg=1.5\times 10^{13}$\,km, Messenger data go a little above 3 m higher than the reference value, but decrease then as $\lg$ decreases, while Mars standard deviation does exactly the opposite --- indicating a compensating mechanism between the two sets of standard deviations, whose controlling parameters are indeed highly correlated. Of course, the compensating mechanism is actually across the whole set of residuals and depends on both the weights attributed to the different data and to the correlations between various parameters. 
In the end, only residuals deduced from Cassini show a significant (and monotonic) increase as $\lg$ decreases, as one can see in Fig. \ref{fig:sigma}. 

\section{Discussion}

The significant level of correlation between $\lambda_g$ and the Solar System parameters 
 indicates that any signal introduced by $\lambda_g < + \infty$ can in part be re-absorbed during the fit of all the parameters. As a consequence, as explained previously, analyses based on postfit residuals tend to overestimate the constraint on $\lambda_g$. We can illustrate this further with the following example.

\begin{figure}
    \includegraphics[scale=0.5]{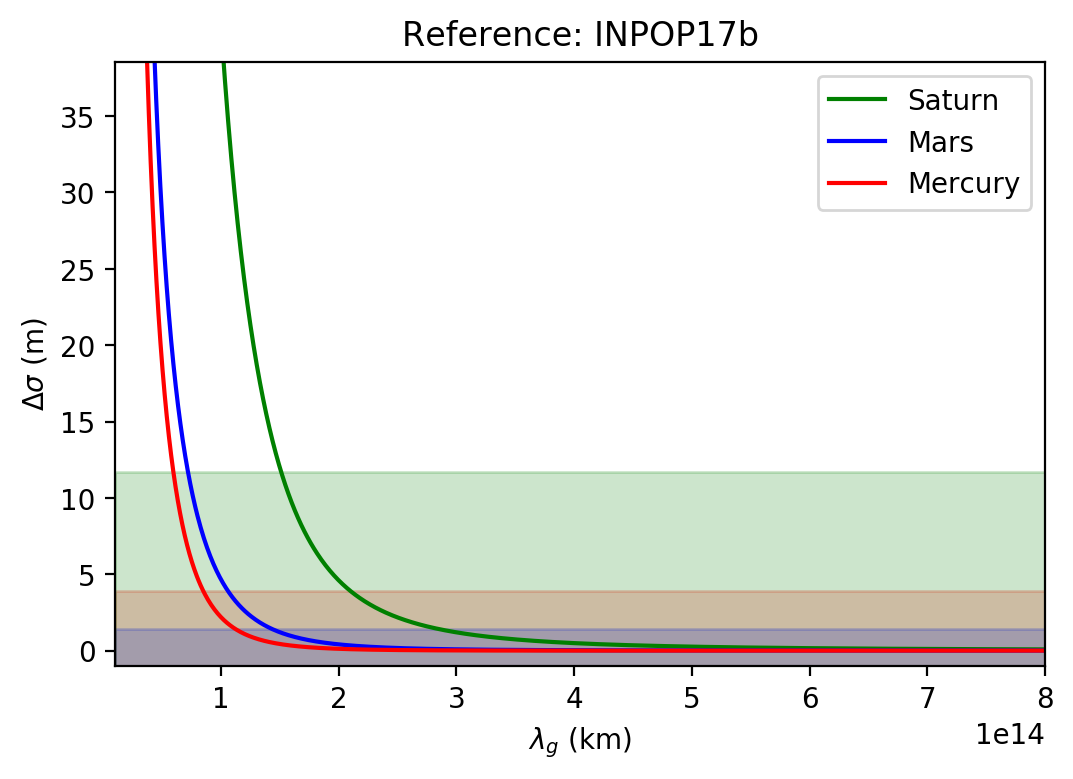}
	\caption{Plot of the standard deviations variations with respect to $\lg$ without adjusting the Solar System parameters. As expected, $\lambda_g$ seems to have a much more important impact compared to the case in Fig. \ref{fig:sigma}, where one has adjusted the Solar System parameters after adding the $\lambda_g$ parameter. [Note that the scale of the x-axis here is different with respect to Fig. \ref{fig:sigma}].}
        \label{fig:sigma_unfit}
\end{figure}

\begin{figure}
        \includegraphics[scale=0.5]{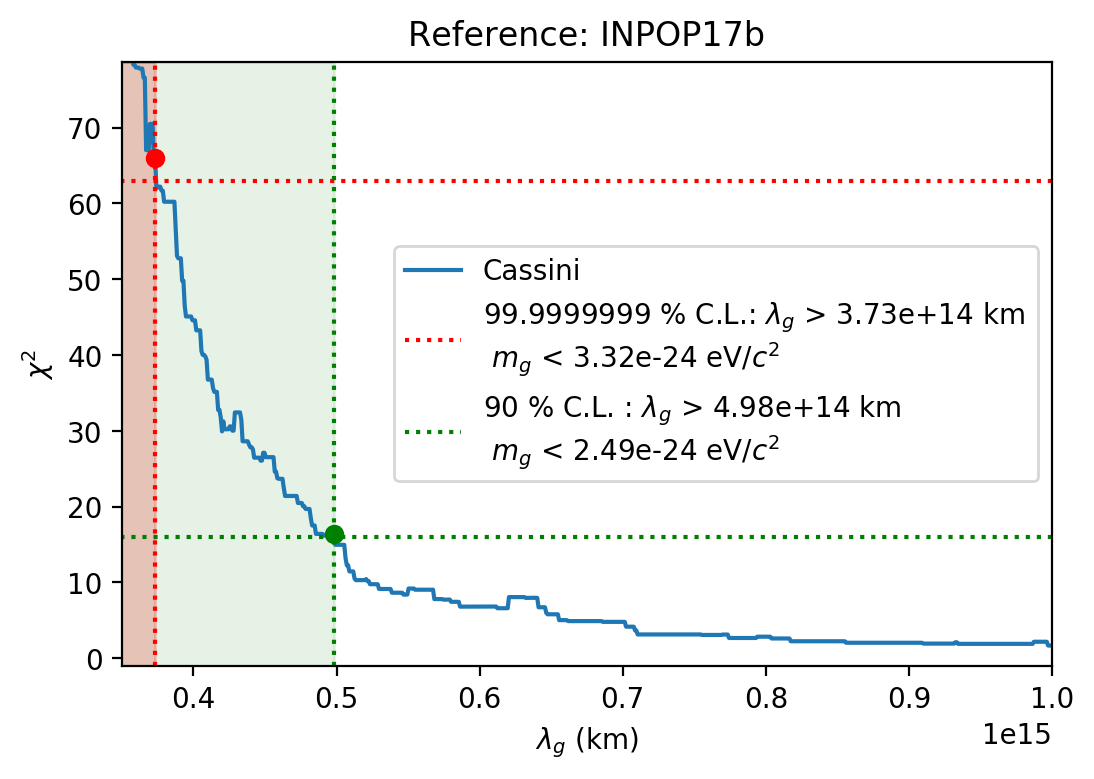}
        \caption{Plot of $\chi^2(\lg)$, and the constraints one would have deduced for $\lg$ without re-adjusting the Solar System parameters. It would lead to a spurious improvement of a factor \colp{27.2} and \colp{22.4} for the 90\% and 99.9999999\% C.L. bounds respectively, compared to the global fit analysis.}
        \label{fig:chi2_unfit}
\end{figure}

We look at the standard deviations of residuals with respect to $\lambda_g$ without re-adjusting all the Solar System parameters --- which effectively corresponds to do a sort of postfit analysis, in the sense that one considers Solar System parameters to be given beforehand by previous analyses. This gives an indication of the amplitude of the effect of $\lambda_g < + \infty$ on the ephemeris, when the Solar System parameters are (wrongly) considered to be known beforehand. The standard deviations for Saturn Cassini residuals are shown in Fig. \ref{fig:sigma_unfit}. If one applies the same statistical analysis on those residuals --- instead of on the residuals obtained after adjusting the Solar System parameters as in Fig. \colp{1} of the main part of the manuscript --- one would mistakenly get at an improvement of more than a factor \colp{20} on the constraint of $\lambda_g$, as it can be seen in Fig. \ref{fig:chi2_unfit} (the postfit analysis gives $\lg\le 4.98\times10^{14}$\,km instead of $1.83\times 10^{13}$\,km, for the 90\% C.L. bound).

